# Anomaly-Based Intrusion Detection System for Cyber-Physical System Security

Riccardo Colelli          Filippo Magri [y]          Stefano Panzieri [z]          Federica Pascucci [x]

## Abstract

Over the past decade, industrial control systems have experienced a massive integration with information technologies. Industrial networks have undergone numerous technical transformations to protect operational and production processes, leading today to a new industrial revolution. Information Technology tools are not able to guarantee confidentiality, integrity and availability in the industrial domain, therefore it is of paramount importance to understand the interaction of the physical components with the networks. For this reason, usually, the industrial control systems are an example of Cyber-Physical Systems (CPS). This paper aims to provide a tool for the detection of cyber attacks in cyber-physical systems. This method is based on Machine Learning to increase the security of the system. Through the analysis of the values assumed by Machine Learning it is possible to evaluate the classification performance of the three models. The model obtained using the training set, allows to classify a sample of anomalous behavior and a sample that is related to normal behavior. The attack identification is implemented in water tank system, and the identification approach using Machine Learning aims to avoid dangerous states, such as the overflow of a tank. The results are promising, demonstrating its effectiveness.

## 1  INTRODUCTION

Factors such as technological development, the application of Internet of Thing (IoT) to the industrial world and the birth of all the applications characterizing the industry 4.0, have meant that the structure of industrial control systems (ICSs), previously closed networks, has opened up to the network that evolves towards Networked Control Systems (NCS).

Activities such as monitoring and control are managed through communication channels in Supervision Control and Data Acquisition (SCADA) systems, over vast geographical areas. Hence this technological evolution has led to considerable performance advantages, on the other hand, it has broadened the range of potential threats to which these systems are exposed. In fact, actor such as hostile groups or discontented employees now represent sources of possible threats [16]. Furthermore, if we consider the fact that the use of such systems is often aimed at monitoring and controlling a critical infrastructure, the issue of their safety acquires even greater relevance as any problems may have repercussions for public safety [18]. Based on this, it is not surprising the growing interest in research towards Cyber-Physical Security which constitutes a central issue in the management of modern ICS.

With regard to the protection and security of data and Information Technology (IT) resources, there are three fundamental principles on which attention must be focused: confidentiality, integrity and availability (CIA) [4]. Unlike the traditional IT vision, the order of importance of the features described above follows the CIA paradigm, when it comes to security in the industrial sector, the paradigm is reversed in AIC [7].

Riccardo Colelli is with Department of Engineering, University of Roma Tre, 00146 Rome, Italy, riccardo.colelli@uniroma3.it

[y]Filippo Magri is with Department of Engineering, University of Roma Tre, 00146 Rome, Italy, fil.magri@stud.uniroma3.it

[z]Stefano Panzieri is with Department of Engineering, University of Roma Tre, 00146 Rome, Italy, stefano.panzieri@uniroma3.it

[x]Federica Pascucci is with Department of Engineering, University of Roma Tre, 00146 Rome, Italy, federica.pascucci@uniroma3.it

Furthermore, protect data in Operational Technology (OT) is crucial for the plants and resources given that the unavailability of some resources could cause serious damage to people and to the environment. Our work is motivated by cyber attacks involving different types of critical infrastructures. One of them is Stuxnet attack [1]. Through Stuxnet malware, the Programmable Logic Controllers (PLCs) used to control centrifuges in a nuclear plant were tampered and the resulting malfunctions were prevented from being detected. The feature that immediately struck the experts was the level of sophistication of the software, which emerged after its analysis. The onset of the infection probably took place inside the plant itself, via an infected USB key in the hand of an unsuspecting person. The malware therefore leveraged four Windows vulnerabilities to gain access to the network and then propagate itself in search of PLC control software, then modify its code to damage the system without safety systems and operators become aware of the anomaly.

Securing a computer system involves an architecture consisting of different types of hardware devices and software applications that form a barrier capable of protecting the system from the different types of attacks to which it is exposed. The need arises for a technology capable of monitoring such systems and consequently identify possible intrusion attempts. This technology is known as the Intrusion Detection System (IDS). Intrusion detection is the process of monitoring and analyzing events that occur in a computer system or network in order to detect intrusions, defined as attempts to compromise the confidentiality, integrity or availability of data or the network [15]. IDS technology in SCADA system needs to be extended considering the data of the controlled process [19].

The implementation of a defensive solution for ICS is proposed in this paper using a supervised Machine Learning (ML) algorithm. Hence, a classifier is obtained in order to detect anomalies in the system behavior.

## 1.1 Paper Contribution

The methodology used in this paper aims at giving an anomaly detection by using integration between ML and IDS techniques. The methodology is based on the concept of supervised learning in order to obtain a classifier to detect anomalies.

The contributions of the paper are two-fold. Firstly, we apply the concept of ML to an IDS in a cyber-physical system composed of a single water tank with a pump and a valve. Secondly, we apply the detection strategies in a simulated environment. This scenario allows to use a very useful tool that enables research in the context of Cyber-Physical Security and respond to a need of cyber security that seems to be increasingly present. In particular, the simulator based solution demolishes the costs of the hardware and ensuring the work on a system with a physical component.

## 1.2 Organization of the Paper

The rest of the paper is organized as follows. Section 2 describes the identification process for attacks in a CPS and similarities in their detection. We define the used method for Machine Learning in Section 3. Cyber-Physical case study is described in Section 4. Next, we analyzed our detection approach for detecting cyber attacks by describing the attack scenario and the experimental results are presented in Section 5. Conclusions are presented in Section 6.

# 2 RELATED WORKS

The weaknesses and limitations associated with the SCADA systems have been the target of cyber attacks. In according with [7], the connection of physical cyber systems to external networks (e.g. IoT) has exposed these systems to new forms of threats. A typical sabotage in a Cyber-Physical system can take place thanks to the protocols adopted in the industrial control systems. Hence, authors in [12] consider that protocols from ICS are still legacy ones. Those protocol do not consider various cybersecurity requirements such as authentication between devices, integrity of messages, or the confidentiality of the communication channels.

Many different IDS for SCADA systems are developed in order to report anomalous behaviour [13, 3]. The increment of SCADA network traffic makes the monitoring and analysis of data very expansive in order to identify malicious traffic. For this reason, Machine Learning (ML) strategy methods are recently used in IDS to track normal network and system behavior. In [11], a machine learning-based algorithms are proposed to model the network behaviors of SCADA hosts. In [20], network packets up to their application-layer headers are analyzed and a comprehensively model of the network behaviors for SCADA hosts is created by using a neural network. In [21] a cyber-physical testbed is presented by the authors in order to investigate an IDS using ML and artificial neural network models for detect reconnaissance and DoS attacks.

# 3   PROPOSED METHOD

Machine learning term referred to a particular branch of computer science that studies systems and algorithms capable of learning from data synthesis and which can therefore be considered a close relative of artificial intelligence, representing the meeting point between this and data mining. Since, in the implementation of the defensive solution for ICS proposed in this discussion, the use of a supervised learning in machine learning algorithm is envisaged through which to obtain a classifier. This classifier is used in the detection of anomalies in the behavior of the system, the concepts are introduced key at the basis of machine learning techniques. A system based on machine learning is able to acquire knowledge of the context of interest by observing the data that is provided as input and produce output based on the knowledge acquired. By observing the input data, a model is then built through which the algorithm will be able to determine the desired output. Typically, the input is represented by a vector or sample where each element represents a feature, or a characteristic. For instance, assuming input vectors referring to packets passing through a network, the features could be the components defined by the protocol to which it belongs. Defining the characteristics and applications of machine learning in a simple way is not always possible, as machine learning algorithms can be used in a wide variety of applications such as, for example, facial recognition, predictive algorithms of stock market fluctuations or self-driving cars. Therefore, different application areas require different types of algorithms. In this paper Random Forests algorithm is used as described below.

## 3.1   Random Forests

Random Forests (RF) is a supervised learning algorithm that can be used for both classification and regression. The main feature of this type of algorithm is the use of many weak predictors, i.e. single decision trees, where collaboration gives rise to a single strong predictor, i.e. a more accurate model obtained by merging the various decision trees [9]. Therefore, the combination of multiple classifiers allows to improve classification accuracy provided, however, that the classifiers involved are weakly related to each other. Hence, there needs to be little dependence between the models associated with the various classifiers and the training set as there will be a reduction in variance and in the classification error. The steps of the algorithm for creating a forest consisting of k decision trees are:

- creation of weakly correlated k training set starting through the bagging technique, i.e. the partitioning of the starting data set through sampling with replacement;

- training, starting from each training set obtained in step 1, of a decision tree;

- random choice, for each node of each tree obtained in step 2, of some splitting attributes;

- classification of test set data through the choice of the class attribute that most appears among the results obtained from individual trees.

A RF usually uses the Gini Index [5] as a measure for the best split selection, which measures the impurity of a given element with respect to the rest of the classes. For a given training dataset T , the Gini Index can be expressed as

$$\sum \sum_{j \neq 1} (f(C_i T)/|T|)(f(C_j; T)/|T|) \qquad (1)$$

Where $f(C_i; T) = |T|$ is the probability that a selected case belongs to class ($C_i$. Hence, by using a given combination of features, a decision tree is made to grow up to its maximum depth.

## 3.2   Intrusion Detection System

The use of anomaly based IDS presupposes the definition of an operating profile of the system considered normal. Among the possible techniques that can be adopted for this purpose, we analyze in more detail the use of machine learning techniques as they are the basis of the work implemented. Techniques based on machine learning have as their objective the construction of models capable of classifying the nature of the instances to be analyzed. The models are built during the training phase, through the use of a data-set partition called training-set. The data is usually tagged but there are also solutions that involve unlabeled data. A fundamental feature of this type of systems is to increase knowledge about the system through the acquisition of new information and thus improve the ability to correctly classify the events analyzed. It should be emphasized that machine learning-based applications require a considerable amount of space-time resources, first of all in the training phase.

# 4   CASE STUDY

One of the main difficulties encountered in the implementation of defensive solutions that aim to counter cyber attacks against ICS systems are the limited possibility of validating their effectiveness in conditions that are as realistic as possible. The difficulties derive mainly from the critically of the application context within which these systems normally operate which often does not allow the conduct of the experiments necessary for the development of valid countermeasures. Therefore, it is essential to use auxiliary tools, such as testbeds or specific simulation environments, through which it is possible to subject the systems analyzed to the various attack scenarios, in order to refine the defensive measures through which one wants to limit their impact. A very useful tool in this context is MiniCPS [2] framework, by extending the functionality of Mininet [6], that allows to simulate a vast typology of Cyber-Physical System with a high level of detail. Hence, by using a CPS simulator is possible to provide an application example of a proper use of MiniCPS aimed at the implementation of defensive solutions for cyber-physical systems. The solutions that we implement foresee the development of a host and anomaly-based Intrusion Detection System that uses machine learning mechanisms with supervised learning for the definition of system behavior profiles to be used in the detection of anomalies. All the steps that allowed the development of these defensive solutions are then shown in detail, starting from the implementation of the software tools through which it was possible to conduct the simulation of the system as a whole, to conclude with the simulation of data modification attacks and their detection. The process considered does not have many variables to control but it is functional for the application of the ML. In particoular, the system consists of a reservoir of water which is iteratively filled and emptied, as shown in Fig. 1. The water level inside the tank is monitored through the use of a level sensor and controlled by switching an on/off pump, which determines its inlet flow, and the opening or closing of a valve, which instead determines the outflow. Both flows are in turn monitored through the use of two additional sensors. Regarding the tank level, four thresholds are considered:

- LL threshold: it is the threshold below which the tank level must not go;

o L threshold: when the tank level is below the threshold L, the pump that regulates the inlet flow is turned on and the valve that regulates the outflow is closed, causing the tank to be filled;

o H threshold: when the tank level is above the threshold H, the pump that regulates the inlet flow is switched off and the valve that regulates the outflow is opened, causing the tank to empty;

- HH threshold: it is the threshold beyond which the tank level must not go.

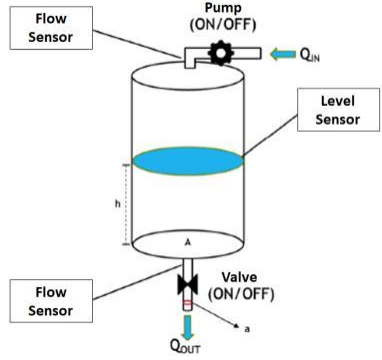

Figure 1: Single water tank system.

The equations of the system are as follow:

$$Q_{IN} = \alpha \cdot P \quad\quad\quad\quad (2)$$

$$Q_{OUT} = \beta \cdot \alpha \sqrt{2gh} \quad\quad (3)$$

$$h = \frac{Q_{in}}{A} - \frac{Q_{out}}{A} \quad\quad (4)$$

Where P is a constant representing the maximum flow associated with the pump. The binary variables $\alpha$ and $\beta$ , both represent the status of the pump, 0 if the pump was off, 1 if it was on. $Q_{IN}$ represents the inflow and $Q_{OUT}$ represents the outflow. $\sqrt{2gh}$ represents the maximum flow associated with the valve, it considers the contribution of gravity (g), the section of the outlet hole (a) and of the tank level (h). A is a constant that represents the section of the tank.

For process control, three control devices are used (i.e. sensors, PLC and actuators). The simulation of the functioning of the three control devices was obtained through the implementation of three different scripts in MiniCPS, as shown in Fig. 2, a control loop allows to read from sensors and to operate actuators.

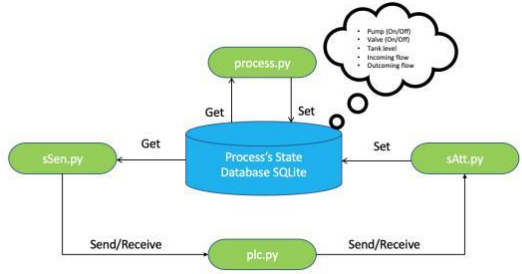

Figure 2: Architecture of the simulated network.

Scikit-learn [10], an open-source library, is used for machine learning. The classifiers for the classification of system anomalies include: Nearest Neighbors Classifier, C-Support Vector Classifier, Decision Tree Classifier, Random Forest Classifier. This latter classifier is considered in the implementation and in the results presented. The performance of the model strongly depends on the choice of hyperparameters. Hyperparameters are adjustable parameters that allow to control the model training process. This process can be expensive. However, scikit-learn library provides an useful tool in identifying the optimal configuration of the hyperparameters of a classifier, namely the GridSearchCV.

The metrics, used for the evaluation of the models obtained from the training are explained below: Accuracy is the indicator that provides the measure of the correct classifications, returned by the classifier, with respect to the total of samples that make up the dataset to be classified (5); Precision is the indicator that provides a measure of a classifier's ability not to label as positive a sample that is actually negative (6); Recall is the indicator that provides a measure of the ability of a classifier to correctly classify all positive instances within the test set (7); F1-Score is a weighted harmonic average of the Precision and Recall metrics, constructed in such a way that the best score is 1 and the worst is 0 (8).

$$\text{Accuracy} = \frac{TP + TN}{TP + TN + FP + FN} \tag{5}$$

$$\text{Precision} = \frac{TP}{TP + FP} \tag{6}$$

$$\text{Recall} = \frac{TP}{TP + FN} \tag{7}$$

$$F1\ Score = \frac{2 x Recall x Precision}{Recall + Precision} \tag{8}$$

## 5 RESULTS

The experimental results are shown in order to validate the proposed defensive solutions in an ICS scenario. The first step in the experimentation phase is about the collection of a sufficient quantity of data by sampling the behavior of the system in different operating conditions. This data collection has allowed the creation of different training sets, through which to train the classifier used in the detection of anomalies, and a test set, through which to analyze the effectiveness of the classification as the samples used in the training phase vary. For this purpose, various simulations of False Data Injection attacks are executed on the system. The attack consists in the modification of the content of the payloads in a Modbus packets sent by the host used for acquiring data from the sensors and received by the PLC responsible for controlling the process and in which the IDS is implemented. Once the data collection phase is completed, the ML algorithm is trained through the use of the various training sets obtained. It is important to obtain an equal number of models to be used in the detection of anomalies. Therefore, the experimentation phase was concluded with the analysis of the performance of the different models obtained in classifying the samples making up the test set built during the data collection phase.

### 5.1 Data collection

In order to build the different training sets for training the classifier to detect anomalies in the data received from the PLC, a first phase of data collection was conducted, as shown in Fig. 3. Also, a test set is created in order to validate the effectiveness of the data. The behavior of the system under different operating conditions was sampled, simulated through the software tools described and analyzed above. The operating conditions considered are:

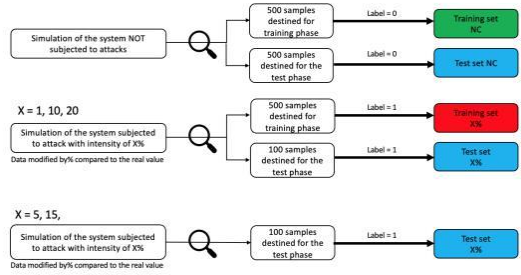

Figure 3: Data collection phase overview.

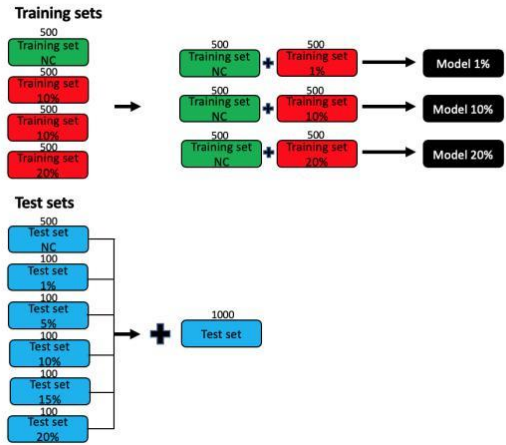

Figure 4: Procedure for the construction of training sets.

Normal operating conditions: a first simulation of the system is implemented, without it being affected by any type of attack, and its behavior was sampled obtaining a set of 1000 samples. This set is divided into two parts of equal size, one intended for training the ML algorithm, i.e. for the composition of different training sets, the other for analyzing the performance of the classifier obtained, i.e. to the composition of the test set.

Abnormal operating conditions: in order to sample the anomalies in the behavior of the system, deriving from the modification of the data in the network, simulations of different attacks are implemented. Attacks differ from the intensity of the modification to which the data was submitted, measured in percentage terms with respect to its real value. The samples collected, for each level of attack intensity, are in some cases divided into two sets. As in the case of the sampling of normal behavior, samples from the training sets and the test set. The attack intensity levels whose samples have been destined for both uses are:

  – attack with 1% modification of the data
  – attack with modification of 10% of the data;
  – attack with modification of 20% of the data;

In all three cases, 500 samples are assigned to training and 100 to the testing phase. The attack intensity levels where samples are intended for the composition of the test set are:

  – attack with modification of 5% of the data;
  – attack with modification of 15% of the data;

In both cases, 100 samples are considered.

All the sets of samples collected, both those intended for training and those intended for the test phase, are labeled (i.e. normal behavior 0 and anomalous behavior 1).

Once the data collection phase is completed, the sets of samples available for training the ML algorithm and to test the classifier returned, as shown in Fig. 4. All three training sets are constructed using the set of 500 samples, relating to the normal behavior of the system, previously collected but, for each of them, a different set of 500 samples is used as a reference for the classification of anomalous behavior. Hence, in all three cases the overall size of the training set is 1000 samples. It is important to underline that the distribution of the samples, used for training, is balanced. The algorithm has an equal number of references, representative of the two different classes, based on which to implement the classification. On the other hand, an unbalanced training set can have important consequences on the classification performance of the ML model, making it lean more towards assigning a sample to the class most populated of the training set.

The construction of the test set, to analyze the classification capacity of the three models obtained as a result of training, uses all the sets of samples accumulated during the data collection phase. The total size of the test set thus obtained is 1000 samples, equally distributed between the 500 samples relating to a normal behavior and the 500 samples relating to an anomalous behavior. The 500 samples related to an anomalous behavior include 5 different levels of intensity of anomaly, each of which represented by 100 samples. Hence, the data on the accuracy of the ML model in classifying the samples of a given test set is representative of the goodness of the classifier. It is necessary that the test set is balanced and that are two in the case in our scenario. In the event that, the classifier is able in the classification of negative samples and is unable to correctly classify the positive samples, a test set composed of a large prevalence of negative samples would lead to an accuracy value very high. Thus, a high number of correct classifications compared to the total of samples does not represent the real performance of the classifier. In addition to the composition of the training set intended for the training of the classifier, also the composition of the test set can greatly influence the values returned by the metrics used to analyze the goodness of a classification model.

## 5.2 Attack scenario

A Man-In-The-Middle (MITM) attack in MiniCPS network is launched by a malicious device.

During normal communication, the attacker starts the MITM attack using the ARP poisoning technique combined with IP forwarding. In this way, all the traffic flows first through the attacker and is later retransmitted to the actual destination, preserving the IP of the original source. It is assumed that the network components can re-establish the connection after a disconnection. Thus, the payload capture begins and the intercepted sensor data information carried by responses from the PLC are stored by the attacker. The payload capture lasts for a prefixed time. Thereby, the acquired packet is stored and the attacker is ready to perform the Data Modification. Data manipulation occurs through the combined use of Scapy [14] and NetFilterQueue [8]. In particular, Scapy is a packet manipulator which can parse several protocols including Modbus. Moreover, we use the NetFilterQueue Python bindings for libnetfilter_queue to redirect all the messages between PLC and sensor/actuators to a handling queue defined on the table of the Linux firewall iptables. Data Modification starts and the queued packets are modified using Scapy. Thereby, adversary changes in the TCP packets only the bytes of the Modbus payload related to the sensor readings and packets reach their original destination. Finally, the MITM is removed, the IP tables are flushed and the original network connection is restored.

## 5.3 Classification results

Once the ML algorithm is trained through the three different training sets described above, we proceeded with the use of the three models obtained for the classification of the samples belonging to the test set in order to analyze the different classification performances. Starting from the observation of the confusion matrices [17], it is possible to observe the influence of a different composition of the training set used in the training of the classifier, on the classification ability of the model. In fact, the use of a set of references which train the classifier to recognize a behavior more distant from that considered normal as anomalous, increases the ability of the model to correctly classify the samples of the test set. On the other hand, it reduces the ability of the same to correctly classify samples that describe a lightly anomalous behavior, as they are closer to the references of normal behavior shown to the classifier during training.

Therefore, we can observe a limited number of False Positives and a substantial number of False Negatives in relation to the use of the third Training set with model 20%, in Fig. 5 (c). Differently, regarding the use of the first training set with model 1%, confusion matrix is shown in Fig. 5 (a), a high number of False Positives is observed due to the use of references very close to normal behavior in the classifier training and a much smaller number of False Negatives is noticed. In the specific case analyzed in the paper, the use of the second training set seems to be the right compromise capable of training the classifier in such a way as to maintain a good ability to classify samples belonging to both types. The results of confusion matrix of model 10% are shown in Fig. 5 (b).

The metrics used for the evaluation of the models obtained following the training of the classifier through the three different training sets are shown in TABLE 1.

Through the analysis of the values assumed by the metrics considered to evaluate the classification performance of the three models in question, we can reach the same considerations made in relation to the observation of confusion matrices. As regards the model obtained through the use of the first training set a low Precision value can be observed that means a reduced ability of the classifier in not classifying as positive (anomalous behavior) a sample that is actually negative (normal behavior). In addition, considering the Recall, a high value indicates the ability of the classifier to correctly classify all positive samples. Moreover, for the model obtained through the use of the third training set, the situation is opposite. In particular, there is Precision that even assumes its maximum value, that is 1, but a significantly reduced Recall value. Using the F1-score to compare the different classification models we can conclude the model trained using the second training set, is the one that offers such an overall classification ability to ensure a compromise between the correct classification of positive and negative samples. Thus, the perfectly balanced nature of the test set makes accuracy a reliable metric to describe the overall quality of the classifier, it can be observed that this last value is higher in correspondence of the second model. However, it is necessary considering that the evaluation of a classification model cannot ignore the context in which it must be used. For instance, if the action consists of detected anomaly sending an alert to the operator, it will be preferable to use a model with a high Recall value, in order to minimize risk factor deriving from the failure to detect an anomaly. Thus, in the case that a sample of anomalous behavior occurs, the classifier has a high probability of classifying it correctly. On the other hand, if the countermeasure resulting from the detection of an anomaly is highly invasive, it will be probably preferable to use a model characterized by a high Precision value.

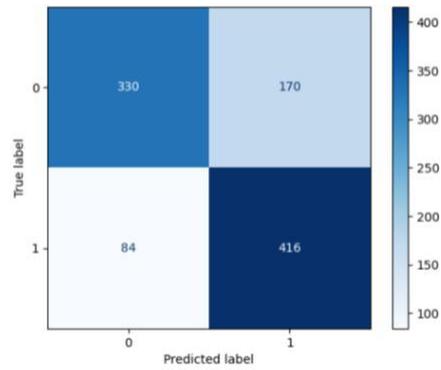

(a) Model 1%

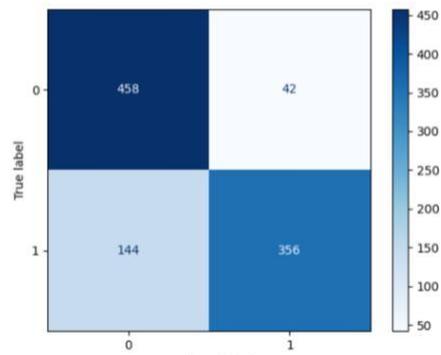

(b) Model 10%

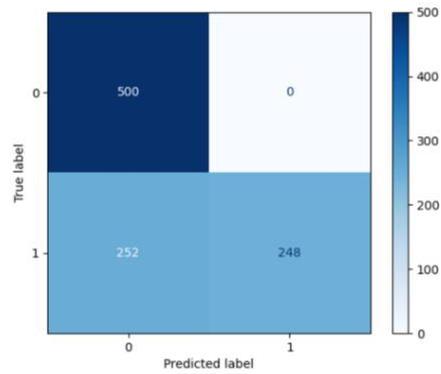

(c) Model 20%

Figure 5: Confusion Matrix.

Table 1: Specification

| Model | Accuracy | Precision | Recall | F1-Score |
|-------|----------|-----------|--------|----------|
| 1% | 74,6 | 0,709 | 0,832 | 0,766 |
| 10% | 81,4 | 0,894 | 0,712 | 0,792 |
| 20% | 74,8 | 1 | 0,496 | 0,663 |

Lastly, the effectiveness of a classifier is related to the intended purpose and the application context.

## 6  CONCLUSIONS

As emerged from the analysis of relevant episodes of attack in CPS, legacy IT defense solutions may not be sufficient to guarantee the protection of ICS systems against increasingly complex and specific attacks with respect to the target system. In the work presented in this paper, it is shown how MiniCPS framework represents a valid solution for create a realistic simulation of the behavior of a Cyber-Physical System also for ML strategies. Therefore, it is possible to implement new defensive solutions aimed at CPSs, validating their effectiveness through a series of experiments that would otherwise be difficult to conduct in real systems. The use of a CPS simulator also allows to integrate your code in a simple and effective way in order to further extend the available functions. This makes process simulation a very useful tool that can help research in the context of Cyber-Physical Security and respond to a growing need. In this paper, we present how the use of Machine Learning in the context of the development of anomaly-based IDS can constitute a valid solution that, together with other legacy approaches, can significantly increase the security of CPSs. We emulate an attack scenario over a single water tank system with a relative control for filling and emptying. We use a machine learning algorithm with supervised learning (Random Forest) for the classification of system behavior in the implementation of an anomaly-based Intrusion Detection System. Training and data set are composed considering different intensity level attack, obtained from a different percentage of data modification. In particular, we have made evident how a different composition of the training set can affect the classification capabilities of the algorithm so that the calculator is optimal with respect to the necessary use.

## ACKNOWLEDGMENT

The current work has in parts been supported by the EU projects RESISTO (Grant No. 786409) on cyber-physical security of telecommunication critical infrastructure.

## References

[1] Antiy Labs. Report on the worm stuxnet's attack, October 2010.

[2] Daniele Antonioli and Nils Ole Tippenhauer. Minicps: A toolkit for security research on cps networks. In Proceedings of the First ACM workshop on cyber-physical systems-security and/or privacy, pages 91–100, 2015.

[3] Amaury Beaudet, Franck Sicard, Cédric Escudero, and Éric Zamaï. Process-aware model-based intrusion detection system on filtering approach: Further investigations. In 2020 IEEE International Conference on Industrial Technology (ICIT), pages 310–315. IEEE, 2020.

[4] Matthew A Bishop. The art and science of computer security. 2002.

[5] Leo Breiman, Jerome Friedman, Charles J Stone, and Richard A Olshen. Classification and regression trees. CRC press, 1984.

[6] Bob Lantz, Brandon Heller, and Nick McKeown. A network in a laptop: rapid prototyping for software-defined networks. In Proceedings of the 9th ACM SIGCOMM Workshop on Hot Topics in Networks, pages 1–6, 2010.

[7] Leandros A Maglaras, Ki-Hyung Kim, Helge Janicke, Mohamed Amine Ferrag, Stylianos Rallis, Pavlina Fragkou, Athanasios Maglaras, and Tiago J Cruz. Cyber security of critical infrastructures. Ict Express, 4(1):42–45, 2018.

[8] NetfilterQueue. Python bindings for libnetfilter_queue.

[9] Mahesh Pal. Random forest classifier for remote sensing classification. International journal of remote sensing, 26(1):217–222, 2005.

[10] Fabian Pedregosa, Gaël Varoquaux, Alexandre Gramfort, Vincent Michel, Bertrand Thirion, Olivier Grisel, Mathieu Blondel, Peter Prettenhofer, Ron Weiss, Vincent Dubourg, et al. Scikit-learn: Machine learning in python. the Journal of machine Learning research, 12:2825–2830, 2011.

[11] Rocio Lopez Perez, Florian Adamsky, Ridha Soua, and Thomas Engel. Machine learning for reliable network attack detection in scada systems. In 2018 17th IEEE International Conference On Trust, Security And Privacy In Computing And Communications/12th IEEE International Conference On Big Data Science And Engineering (TrustCom/BigDataSE), pages 633–638. IEEE, 2018.

[12] Juan Enrique Rubio, Cristina Alcaraz, Rodrigo Roman, and Javier Lopez. Current cyber-defense trends in industrial control systems. Computers & Security, 87:101561, 2019.

[13] Naoum Sayegh, Imad H Elhajj, Ayman Kayssi, and Ali Chehab. Scada intrusion detection system based on temporal behavior of frequent patterns. In MELECON 2014-2014 17th IEEE Mediterranean Electrotechnical Conference, pages 432–438. IEEE, 2014.

[14] Scapy. Packet crafting for python2 and python3.

[15] Karen Scarfone and Peter Mell. Guide to intrusion detection and prevention systems (idps). Technical report, National Institute of Standards and Technology, 2012.

[16] Keith Stouffer, Joe Falco, and Karen Scarfone. Guide to industrial control systems (ics) security. NIST special publication, 800(82):16–16, 2011.

[17] James T Townsend. Theoretical analysis of an alphabetic confusion matrix. Perception & Psychophysics, 9(1):40–50, 1971.

[18] David Umsonst and Henrik Sandberg. Anomaly detector metrics for sensor data attacks in control systems. In 2018 Annual American Control Conference (ACC), pages 153–158. IEEE, 2018.

[19] Jared Verba and Michael Milvich. Idaho national laboratory supervisory control and data acquisition intrusion detection system (scada ids). In 2008 IEEE Conference on Technologies for Homeland Security, pages 469–473. IEEE, 2008.

[20] Huan Yang, Liang Cheng, and Mooi Choo Chuah. Deep-learning-based network intrusion detection for scada systems. In 2019 IEEE Conference on Communications and Network Security (CNS), pages 1–7. IEEE, 2019.

[21] Maede Zolanvari, Marcio A Teixeira, Lav Gupta, Khaled M Khan, and Raj Jain. Machine learning-based network vulnerability analysis of industrial internet of things. IEEE Internet of Things Journal, 6(4):6822–6834, 2019.